\documentclass[10pt,conference]{IEEEtran}
\usepackage[english]{babel} 
\newcommand{\He}{\operatorname{H}}
\newcommand{\Tr}{\operatorname{T}}

\newcommand{\matt}[1]{\mathbf{#1}}
\newcommand{\vect}[1]{\mathbf{#1}}

\newcommand{\jim}{\mathrm{j}\,}

\usepackage{amsmath}
\usepackage{amssymb}
\usepackage{epsfig}
\usepackage{psfrag}
\usepackage{tikz}
\usetikzlibrary{arrows,snakes,shapes,positioning,arrows,decorations.markings,calc}
\usetikzlibrary{dsp,chains}

\DeclareMathOperator*{\diag}{\textrm{diag}}

\begin{document}
\selectlanguage{english}
\title{{Spectral shaping with low resolution signals}}
\author{{Hela Jedda, Amine Mezghani and Josef A. Nossek}
\authorblockA{\\Institute for Circuit Theory and Signal Processing\\ Technische Universit\"at M\"unchen, 80290 Munich, Germany\\
E-Mail: \{hela.jedda, amine.mezghani, josef.a.nossek\}@tum.de}\vspace{-0.8cm}}
\maketitle 
\tikzset{DSP lines/.style={help lines,very thick,color=black}}
\tikzset{line_arrow/.style={help lines,very thick,color=black,->,-angle 90}}
\tikzset{filter/.style={rectangle,inner sep=0pt,minimum height=0.8cm,minimum width=1.1cm,draw=black,very thick}}
\tikzset{delay/.style={rectangle,inner sep=0pt,minimum size=1cm,draw=black,very thick}}
\tikzset{downsampling/.style={rectangle,inner sep=0pt,minimum height=0.8cm,minimum width=0.7cm,draw=black,very thick}}
\tikzset{upsampling/.style={rectangle,inner sep=0pt,minimum height=0.8cm,minimum width=0.7cm,draw=black,very thick}}
\tikzset{empty_node/.style={inner sep=0pt,minimum size=0cm}}
\tikzset{connection/.style={circle,draw=black,fill=black,inner sep=0pt,minimum size=2mm}}
\tikzset{coefficient/.style={isosceles triangle,draw=black,very thick,inner sep=0pt,minimum size=.7cm}}
\tikzset{source/.style={semicircle,minimum size=.5cm,draw=black,very thick,shape border rotate=270}}
\tikzset{adder/.style={circle,minimum size=.25cm,inner sep=0pt,draw=black,very thick}}
\tikzset{multiplier/.style={circle,minimum size=.25cm,inner sep=0pt,draw=black,very thick}}
\tikzset{double_arrow/.style={double distance=5pt,thick,shorten >= 6pt,decoration={markings,mark=at position 1 with {\arrow[scale=.6,>=angle 90]{>}}},postaction={decorate}}}
\begin{abstract}
We aim at investigating the impact of low resolution digital-to-analog converters (DACs) at the transmitter and low resolution analog-to-digital converters (ADCs) at the receiver on the required bandwidth and the required signal-to-noise ratio (SNR). In particular, we consider the extreme case of only 1-bit resolution (with oversampling), where we propose a single carrier system architecture for minimizing the spectral occupation and the required SNR of 1-bit signals. In addition, the receiver is optimized to take into account the effects of quantization at both ends. Through simulations, we show that despite of the coarse quantization, sufficient spectral confinement is still achievable. 
\end{abstract}

\section{Introduction}The ever increasing requirements for higher capacity radio networks makes it important to be able to provide very easily installed, high capacity, low power RF-node solutions. The deployment of a large number of antennas and the access to more bandwidth are considered as key enablers to achieve higher data rates in future wireless networks \cite{Ngo13,Ran14}. To this end, mm-wave communication is a very attractive technology, which provides access to a vast amount of spectrum in the mm-wave band \cite{Ran14,Roh14,swindlehurst}. Due to the smaller wavelength, more antennas can be packed in the same volume as compared to current microwave communication systems, and hence, we come upon arrays with large number of antennas at the base stations, i.e. the so-called massive MIMO \cite{Mar10}.  However, there are several issues that need to be addressed for the implementation of massive MIMO in practice. One of the major limiting factors for the implementation of massive MIMO are the complexity issues and the energy consumption due to the large amount of antennas \cite{Bjo14}. In particular, high resolution ADCs/DACs with high sampling rate (several Gsps) will require extremely high  power consumption and cost, which makes their usage for massive MIMO  with individual ADC/DAC for each antenna unbearable. A potential solution to handle this bottleneck is the utilization of low cost and low power RF components \cite{Bjo13}, such as low resolution, for instance 1-bit, ADCs and DACs together with low complexity modulation (e.g. QPSK). In fact, the analysis of the quantization process has gained a lot of attention in the academic research due to its practical relevance  \cite{mezghani_isit2010,Hea14, heath1, heath2, singh, wang, desset}. The front-end structure of transmitters and receivers equipped with a larger number of antennas can be significantly simplified by employing such low resolution devices. In addition, the quantization loss can be compensated partially by increasing the sampling rate. In the case of a 1-bit ADC/DAC, we point out, that there are significant benefits in terms of relaxation for the automatic gain control and requirements on the amplifiers. However, practical concepts of designing low cost and efficient 1-bit transceivers are still an open research area. While signal processing and communications with 1-bit ADCs have gained recently increased interest by the research community, signal preprocessing for 1-bit DACs is still not well established. \\
 	 
In addition to the spatial processing for beamforming and multi-user processing, appropriate waveform designs are needed to confine and separate the transmitted spectra. Two different approaches, namely multicarrier and single carrier transmission, are commonly studied and considered in the literature in the context of MIMO transmission.   
   On the one hand, multicarrier systems provide a natural solution to the frequency selectivity of propagation channels, which simplifies their equalization. This is very attractive for ultra-wide band channels which  are subject to multipath propagation.  Filter bank based multicarrier systems (FBMC) offer a number of benefits over conventional orthogonal frequency division multiplexing (OFDM) with cyclic prefix (CP). One benefit is the improved spectral efficiency by not using a redundant CP and by having much better control of out-of-band emission.  This is made possible by using optimized prototype filters and more elaborate equalization concepts  compared to the single-tap per-subcarrier equalizer (Eq) for OFDM with CP. Another advantage is the large flexibility of choosing the number of sub-carriers without affecting the spectral efficiency, which is a very crucial option to cope with the effects of PAPR, hardware limitations (DAC/ADC, HPA,...) and fast channel variations. For lower frequencies, multicarrier techniques may still be interesting, as long as the number of subcarriers is kept rather low, just to implement channel aware scheduling. Alternatively, single carrier transmission with an appropriate waveform generated by a pulse shaping filter (implemented in the frequency domain or time domain) might be also an attractive solution for higher frequencies due the prominent advantage of lower PAPR compared to multicarrier systems at the cost of additional processing complexity.  Nevertheless, both approaches, especially the multicarrier approach are likely to be affected heavily by a low resolution DAC. In fact, under severe quantization , single carrier transmission seems to be more appropriate. Therefore, we aim here at considering the impact of one-bit oversampled DACs on the spectral shape and bandwidth occupation as well as the uncoded bit error ratio (BER) in single carrier. The main idea is to construct and analyze a modulation and waveform design technique having compact spectra and being compatible  with low complexity transmitter and receiver implementations (e.g. low resolution ADC/DAC). In addition a receiver design taking into account the effects of quantization will be also presented. 
   
Simulation results show that good performance can be obtained when optimizing the parameters of the pulse shaper (PS) (like roll-off factor and fractional delay) as well as the receiving filter taking into account the effects of the quantization. In particular, we observed a kind of trade-off between the bandwidth occupation and the required SNR to achieve a certain BER. 

Notation: Bold letters indicate vectors and matrices, non-bold letters express scalars. The operators $(.)^{*}$, $(.)^{\Tr}$ and $(.)^{\He}$ stand for complex conjugation, the transposition and the Hermitian transposition, respectively. The $n \times n$ identity matrix is denoted by $\mathbf{I}_{n}$. $\matt{ C_{xy}}$ denotes the covariance matrix between the vectors $\vect x$ and $\vect y$. $\matt {K_{xy}}$ is defined as the inverse square-root of the diagonal matrix containing only the diagonal elements of the covariance matrix $\matt {C_{xy}}$, i.e. $\matt {K_{xy}}= \diag(\matt {C_{xy}})^{-1/2}$.

\section{System Model}
\label{sec:system_model}
\begin{figure}[h]
\centering
\resizebox{8cm}{!} {%
\begin{tikzpicture}

\node (in){};
\node[upsampling] (upsampling) [right=of in]{$\uparrow {\ell}_u$  };
\node[filter] (pulseshaper) [right=of upsampling]  {PS};
\node (n1)[right=of pulseshaper] [xshift=-0.9cm, yshift=0.5cm] {};
\node[filter] (dac) [right=of pulseshaper] {1-bit DAC};
\node[filter] (lpf) [right=of dac]   {LPF};
\node[filter]  (adder2)  [right=of lpf][xshift=-0.4cm,yshift=-0.8cm, rotate=-90] {AWGN};
\node [filter] (receiver) [below=of lpf][yshift=-1cm]{LPF};
\node [filter] (adc) [left=of receiver]   {1-bit ADC};

\node [filter] (eq) [left=of adc]  {Eq};
\node[downsampling] (downsampling) [left=of eq]{$\downarrow {\ell}_d$};
\node (out) [left=of downsampling]{};

\draw[DSP lines] [-stealth] (in.east) -- (upsampling.west) node[pos=0.5,above]{$s[m]$};;
\draw[DSP lines] [-stealth] (upsampling.east) -- (pulseshaper.west) node[pos=0.5,above]{$s_u[n]$};

\draw[DSP lines] [-stealth] (pulseshaper.east) -- (dac.west) node[pos=0.5,above]{$y[n]$};;
\draw[DSP lines] [-stealth] (dac.east) -- (lpf.west) node[pos=0.5,above]{$y_Q(t)$};;
\draw[DSP lines] [-stealth] (lpf.east) -| (adder2.west) node[pos=0.5,above]{$y_t(t)$};;
\draw[DSP lines] [-stealth] (adder2.east) |- (receiver.east);
\draw[DSP lines] [-stealth] (receiver.west) -- (adc.east)node[pos=0.5,above]{$x(t)$};;
\draw[DSP lines] [-stealth] (adc.west) -- (eq.east) node[pos=0.5,above]{$x_Q[n]$};
\draw[DSP lines] [-stealth] (eq.west) -- (downsampling.east);
\draw[DSP lines] [-stealth] (downsampling.west) -- (out)node[pos=0.5,above]{$\hat s[m]$};
\end{tikzpicture}%
}
\caption{Single carrier communication system}
\label{fig:sys_model}
\end{figure}
Consider the block diagram of a communication system transmitting QPSK symbols generated at a symbol period of $T_s$ depicted in Fig. \ref{fig:sys_model}. In the transmitter, the QPSK symbols $s[m]$ from $\mathcal{CN}\left(0,\sigma_s^2\right)$ are shaped by a PS in the digital domain at a sample distance of $T = T_s/{\ell}_u$. The resulting baseband signal is converted to the analog domain, utilizing two 1-bit DACs for the inphase and quadrature components. The inphase and quadrature parts of the input signal $y[n]$ are quantized and mapped to either $+1$ or $-1$ depending on their signs and then converted to the continuous-time domain by sample and hold method. The low-pass filter (LPF) removes the DACs' alias spectra. The output signal $y_t(t)$ of transmit power $P_{\text{T}}$ propagates through an AWGN channel between the transmitter and receiver and gets disturbed by some Gaussian distributed noise of power spectral density $N_0$. The signal-to-noise ratio (SNR) is then defined as $\text{SNR} = \alpha \frac{P_{\text{T}}}{\sigma^2_n}$, where $\alpha$ represents the constant channel power gain and $\sigma^2_n = N_0/T_s$. The signal captured by the receiver antenna filtered by a LPF to limit the noise bandwidth. The baseband signal is converted into discrete-time domain and then is quantized using two 1-bit ADCs for the inphase and quadrature components. After the ADCs, the Eq filters the received signal to recover the desired signal $s$. Finally, the signal gets downsampled to obtain the symbols $\hat s [m]$ at the symbol rate. In this work, we investigate the design of the PS and the Eq in the time domain when using 1-bit DAC and ADC.

\section{Design in the time domain}
\subsection{Pulse shaper design}
\begin{figure}
\centering  
\psfrag{continuous-time}[][]{\tiny$h_{RRC}(t)$}
\psfrag{discrete-time,deltan0}[][]{\tiny$h_{RRC}[n]\!,\!\Delta n\!\!=\!\!0$}
\psfrag{discrete-time,deltan05}[][]{\tiny$h_{RRC}[n]\!,\!\Delta n\!\!=\!\!0.5$}
\psfrag{t}[][]{$t/T_s$}
\epsfig{file= 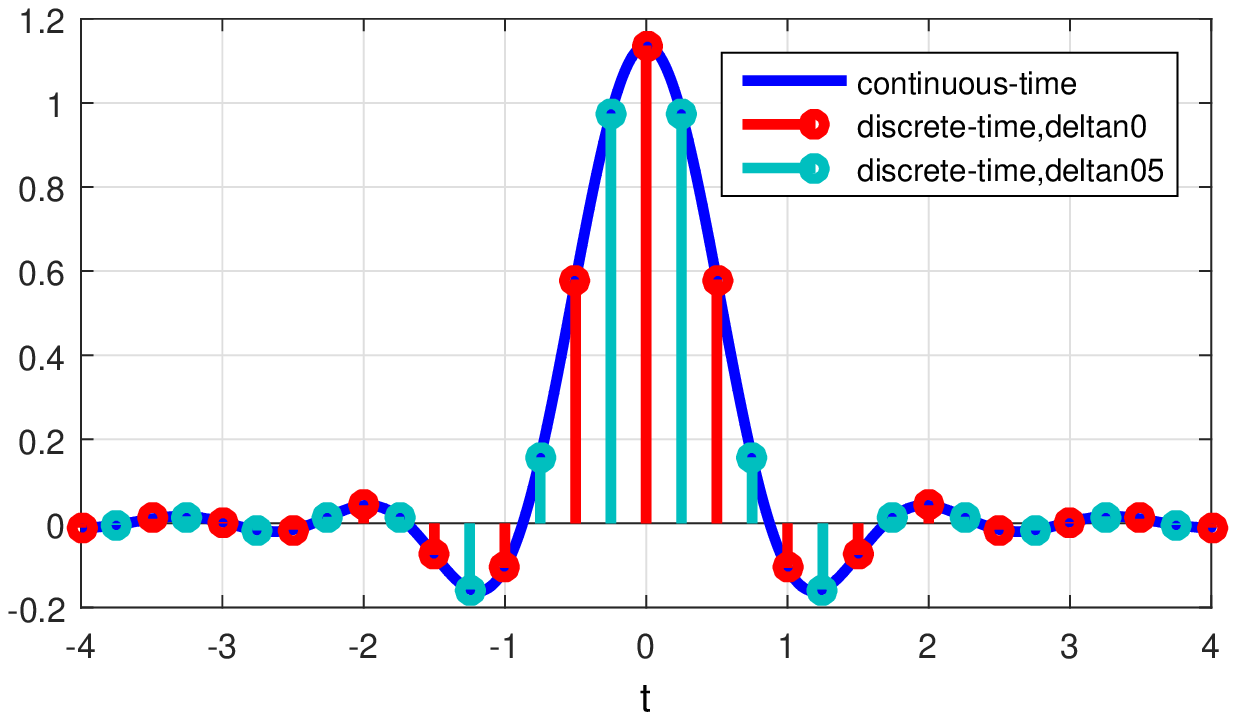, width = \columnwidth}
\caption{Different discrete-time RRC impulse responses for different fractional delay values $\Delta n$: $\rho=0.5$.}
\label{fig:rrc_td_delta_n}
\end{figure}

The commonly used PS is the root-raised cosine (RRC). The continuous-time RRC impulse response is given by
\begin{align*}
h_{RRC}(t) = \frac{4\rho \cos(\pi \frac{t}{T_s}(1+\rho))+\sin(\pi \frac{t}{T_s}(1-\rho))}{\pi \frac{t}{T_s}(1-(4\rho\frac{t}{T_s})^2)},
\end{align*}
where $\rho$ denotes the roll-off factor. Since the PS is processed in digital domain we need an expression for the RRC in discrete-time domain. We get it by sampling the continuous impulse response as follows
\begin{align*}
h_{RRC}[n] \!\!= \!\! \sum^{+ \infty}_{n=- \infty} h_{RRC}(t) \delta(t-(n+\Delta n) T),  n=0, ..., L_{ps}, 
\end{align*}
where $\delta(t)$ is the unit impulse function and $0 \leq \Delta n < 1$. By varying the fractional delay $\Delta n$ we employ the sampling operation at different time instances so that we get a different discrete-time RRC impulse response for each $\Delta n$ as depicted in Fig. \ref{fig:rrc_td_delta_n}. In addition, the 1-bit DAC is a non-linear function, since its output is simply the sign of the input. And here the question arises: which discrete-time RRC impulse response, i.e. which fractional delay $\Delta n$, does lead to the best performance of the non-linear communication system in terms of the spectrum confinement at the transmitter output and the required SNR at the uncoded BER of $10^{-3}$?

\subsection{Equalizer design}
The time-domain Eq is denoted by $\vect w= \left[w[0], ..., w[L_{eq}-1] \right] $. For simplicity we choose the MMSE Eq that is expressed as
\begin{align*}
\vect w = \vect{e}_{\nu}^{\Tr}\matt{ C_{x_Qs}}^{\He} \matt{ C_{x_Q}}^{-1},
\end{align*}
where $\nu$ is the delay introduced by the Eq.
Since 1-bit quantization is employed in our system model, we have to consider the resulting distortions in the Eq design. To this end, we introduce the following equations that summarize the distortions of the 1-bit quantization introduced to a Gaussian distributed complex-valued signal. Let us assume that $\vect r$ is the input of the 1-bit quantizer and $\vect {r_Q}$ is the output. We get
\begin{align}
\matt{C_{r_Q}} &= \frac{4}{\pi}\left(  \arcsin\left( \matt{K_r} \Re \{\matt{C_r} \}\matt{K_r} \right) + \jim \arcsin\left( \matt{K_r} \Im \{\matt{C_r} \}\matt{K_r} \right) \right)  \nonumber \\
&\simeq \frac{4}{\pi}\left( \matt{K_r} \matt{C_r} \matt{K_r} + \underbrace{\left( \frac{\pi}{2}-1\right)}_{=c}  \matt I\right) \label{eq:onebit1} \\
\matt{C_{r_Qr}} &= \sqrt{\frac{4}{\pi}} \matt{K_r} \matt{C_r}. \label{eq:onebit2}
\end{align}
The approximation (\ref{eq:onebit1}) holds true for the assumption $\arcsin(x)\simeq x$.

After using (\ref{eq:onebit1}) and (\ref{eq:onebit2}) in the Eq calculations, we end up with the following MMSE Eq expression
\begin{align*}
\vect w &= \vect{e}_{\nu}^{\Tr} \matt{ C_{x_Qs}}^{\He} \matt{ C_{x_Q}}^{-1}  , \nonumber \\
\matt{ C_{x_Q}} &= \frac{4}{\pi}\left(\matt{K_x} \matt {C_x} \matt{K_x} + c \matt I\right), \nonumber \\
\matt{ C_{x}} &=  \frac{4}{\pi} \left( \matt H_a \matt{K_y} \matt{ C_{y}} \matt{K_y}\matt H_a^{\He} +c \matt H_a \matt H_a^{\He}\right)  + {\ell}_d \sigma^2_n \matt \Gamma \matt \Gamma^{\He}, \nonumber \\
\matt{ C_{y}}& =   \matt H_{ps}  \matt {C_{s_u}} \matt H_{ps}^{\He}, \nonumber \\
\matt {\left[ C_{s_u} \right]}_{ij} &=\begin{cases} \sigma^2_s & \text{ if } i=j=k {\ell}_u, k\in \mathcal{N} \\ 0 &\text{ else,} \end{cases}, \nonumber \\
\matt{ C_{x_Qs}} &=  \frac{4}{\pi} \matt{K_x} \matt H_a \matt{K_y} \matt H_{ps} \matt{ C_{s_us}}, \nonumber \\
\matt {\left[ C_{s_us} \right]}_{ij} &=\begin{cases} \sigma^2_s & \text{ if } i=(j-1) {\ell}_u +1 \\ 0 &\text{ else.} \end{cases}
\end{align*}
where $\matt H_a$, $\matt H_{ps}$ and $\matt \Gamma$ represent the convolution matrices of the analog filters, the PS and the LPF at the receiver. Note that the diagonal elements of $\matt{C_y}$ are not equal for $\rho\neq 0$ because of the non-stationarity of the process.
\section{Results}
After providing the expressions of the PS filter and the Eq optimized with respect to the MMSE criterion,  we aim now at evaluating the performance of the designed 1-bit transceiver system in terms of the uncoded BER and bandwidth occupation. For the simulation, we assume for simplicity an AWGN channel, even though a general channel impulse response could be also handled by adapting the Eq design. On the other hand, the LPFs at the transmitter and receiver are implemented as Butterworth filter of fourth order with 3-dB bandwidth corresponding to the data-symbol rate.  The length of the PS and of the Eq is $L_{ps} = 128$ and $L_{eq} = 64$ respectively and  the roll-off factor is taken from the interval $(0,~1]$.  In addition, we select the oversampling factors at both sides $l_u$ and $l_d$ from the set $\{2,4,8\}$. 
The performance measures in this work are
\begin{itemize}
\item the required SNR at uncoded BER of $10^{-3}$
\item and the bandwidth $B_{0.9375}$. The choice of this bandwidth is justified as follows. We consider the signal-to-interference-noise ratio (SINR), wich is defined as SINR$=\alpha \frac{P_{\text{T}}}{\sigma^2_n+\sigma^2_i}$, where $\sigma^2_i$ denotes the interference power of both adjacent channels spaced by $B_{0.9375}$ with the assumption that $\sigma^2_i = 2 \sigma^2_n$. For an SINR value of 10dB we get the 93,75\% bandwidth $B_{0.9375}$:
\begin{align}
\frac{\int_{f_c -B_{\mathrm{PA}}/2}^{f_c+B_{\mathrm{PA}}/2} S(f) df}{\int_{- \infty}^{+ \infty} S(f) df} &= 93.75 \%,
\end{align}
where $S(f)$ is the power sprectral density of the transmitter output signal $y_t(t)$. This bandwidth definition means that $93,75\%$ of the signal power lies inside. The remaining $6.25\%$ lying outside the defined bandwidth $B_{0.9375}$ is then considered as $\sigma^2_i$.
\end{itemize}
\subsection{Optimal fractional delay improves the performance}
\begin{figure}[h]
\centering  
\begin{minipage}{\columnwidth}
\psfrag{rho=0.1}[][]{\footnotesize $\:\:\:\:\:\:\rho=0.1$}
\psfrag{rho=0.3}[][]{\footnotesize$\:\:\:\:\:\:\rho=0.3$}
\psfrag{rho=0.5}[][]{\footnotesize$\:\:\:\:\:\:\rho=0.5$}
\psfrag{rho=0.7}[][]{\footnotesize$\:\:\:\:\:\:\rho=0.7$}
\psfrag{rho=1}[][]{\footnotesize$\:\:\:\:\:\:\rho=1$}
\psfrag{unquantized system, rho=0.1}[][]{\footnotesize \: unquantized,\! $\rho\!=\!0.1$}
\psfrag{l1=l2=2}[][]{\footnotesize${\ell}_u={\ell}_d=2$}
\psfrag{deltan}[][]{\footnotesize$\Delta n$}
\psfrag{Required SNR @BER=10}[][]{\footnotesize Required SNR @BER$=10^{-3}$ in dB}
\epsfig{file= 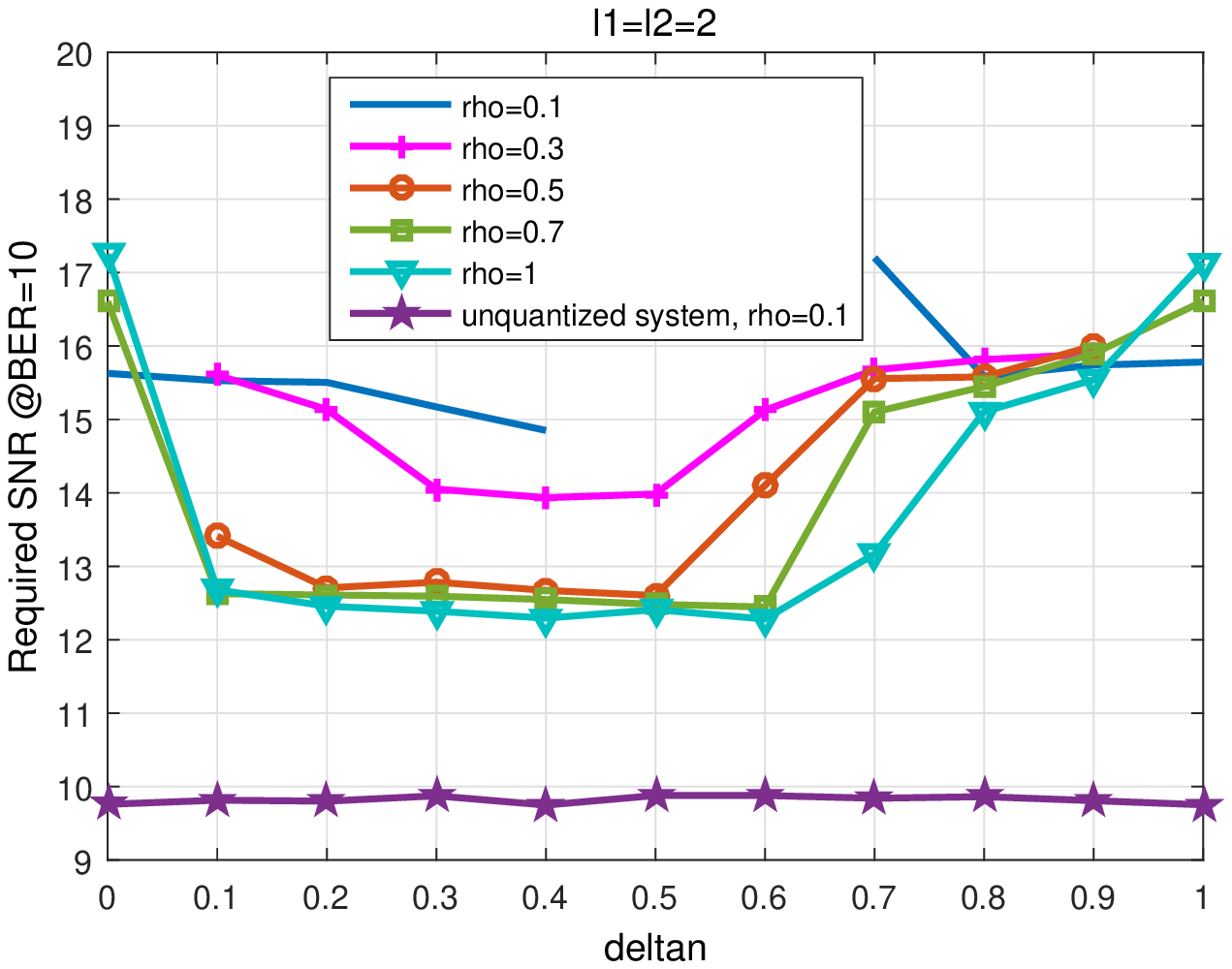, width = \columnwidth}
\end{minipage}
\begin{minipage}{\columnwidth}
\psfrag{rho=0.1}[][]{\footnotesize $\:\:\:\:\:\:\rho=0.1$}
\psfrag{rho=0.3}[][]{\footnotesize$\:\:\:\:\:\:\rho=0.3$}
\psfrag{rho=0.5}[][]{\footnotesize$\:\:\:\:\:\:\rho=0.5$}
\psfrag{rho=0.7}[][]{\footnotesize$\:\:\:\:\:\:\rho=0.7$}
\psfrag{rho=1}[][]{\footnotesize$\:\:\:\:\:\:\rho=1$}
\psfrag{unquantized system, rho=0.1}[][]{\footnotesize \: unquantized,\! $\rho\!=\!0.1$}
\psfrag{l1=l2=2}[][]{\footnotesize${\ell}_u=2$}
\psfrag{deltan}[][]{\footnotesize$\Delta n$}
\psfrag{B}[][]{\footnotesize $B_{0.9375}$ in $1/T_s$}
\epsfig{file= 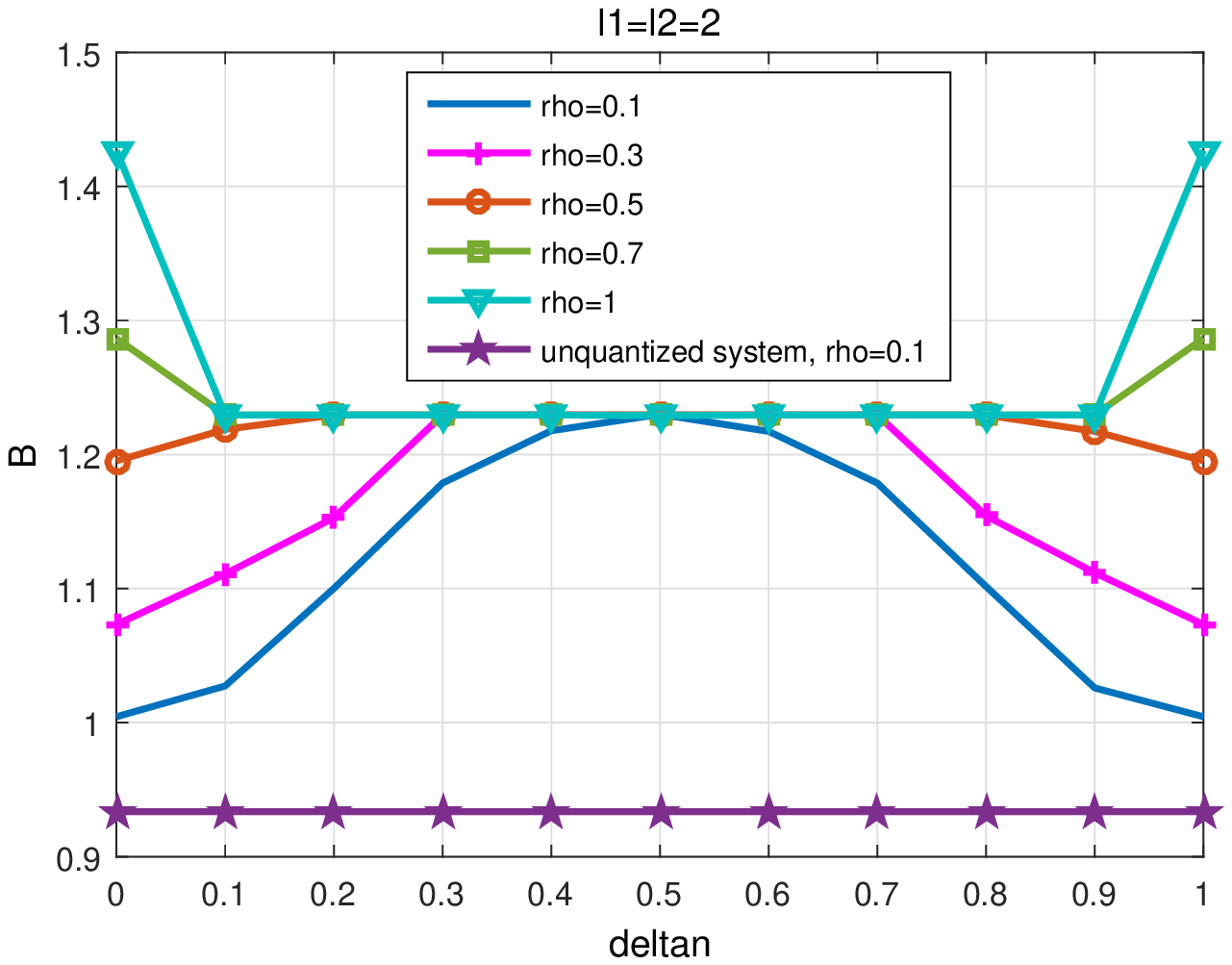, width = \columnwidth}
\end{minipage}
\caption{Required SNR @BER of $10^{-3}$ and $93,75\%$ bandwidth as function of the fractional delay $\Delta n$ for different roll-off factors $\rho$ and for $l_u\!=\!l_d\!=\!2$.}
\label{fig:results_vs_dn_l_2}
\end{figure}

\begin{figure}[h]
\centering  
\psfrag{hrrc time continuous}[][]{ \footnotesize$h_{RRC}(t)$}
\psfrag{hrrc time discrete, deltan=0.5}[][]{\!\!\!\!\!  \footnotesize $h_{RRC}[n]$, $\Delta n\!\!=\!\!0.5$}
\psfrag{t}[][]{$t/T_s$}
\psfrag{Impulse response}[][]{\footnotesize Impulse response}
\epsfig{file= 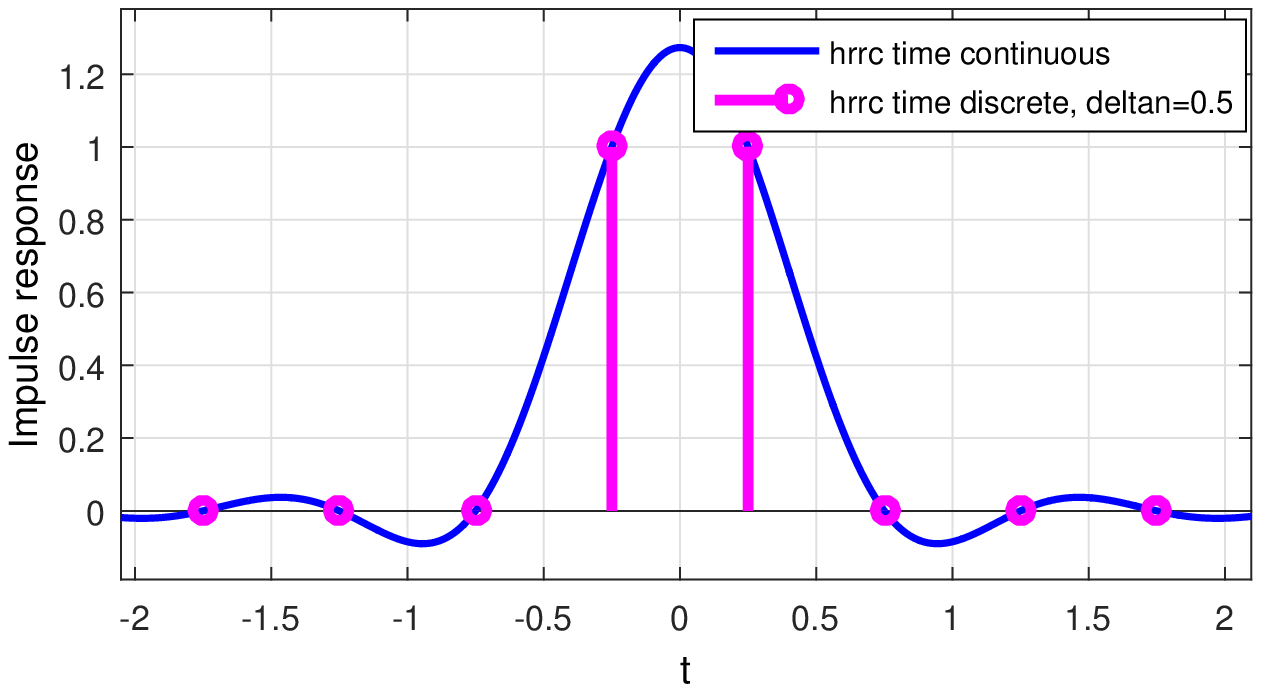, width = \columnwidth}
\caption{Discrete-time RRC impulse response for $\Delta n = 0.5$: $\rho=1$.}
\label{fig:rrc_rho1_delta_n_05}
\end{figure}

Fig.~\ref{fig:results_vs_dn_l_2} shows the performance of the 1-bit system in terms of the required SNR at an uncoded BER of $10^{-3}$ and  $93.75\%$-Bandwidth as function of the fractional delay $\Delta n$ for $\ell_u=\ell_d=2$ and for different roll-off factors $\rho$. One can observe that significant performance improvements are possible by tuning the fractional delay $\Delta n$. For small roll-off values adjusting $\Delta n$ results in a compromise between the required SNR and the bandwidth while for higher roll-off values choosing $\Delta n$ appropriately leads to the best performance for both criteria simultaneously. The minimal required SNR is achieved by the RRC filter with $\rho=1$ and a fractional delay $\Delta n$ between 0.2 and 0.6. We are 2.5dB far from the ideal unquantized case.
If we look closer at the discrete-time impulse repsonse of the RRC filter with roll-off factor $\rho=1$, $\Delta n=0.5$ and with oversampling factor ${\ell}_u=2$, depicted in Fig. \ref{fig:rrc_rho1_delta_n_05}, we observe a two-taps impulse response with unit weighting factors. At the output of this special PS, we get each input symbol, which is a QPSK symbol, repeated twice. This implies that the input of the 1-bit DAC is a sequence of QPSK symbols which makes the non-linearity of the 1-bit DAC ineffective. Thus, we end up with a linear transmitter. We can get rid of the upsampling operation ${\ell}_u$ and the PS in this special case, since their function is in any case realized by the sample and hold function of the DAC. Thus, the removal of the upsampling operation and the PS leads to a reduced system complexity without loosing in the performance.

\subsection{Oversampling improves the performance}
\begin{figure} [h]
\centering  
\begin{minipage}{\columnwidth}
\psfrag{rho=0.1}[][]{\footnotesize $\:\:\:\:\:\:\rho=0.1$}
\psfrag{rho=0.3}[][]{\footnotesize$\:\:\:\:\:\:\rho=0.3$}
\psfrag{rho=0.5}[][]{\footnotesize$\:\:\:\:\:\:\rho=0.5$}
\psfrag{rho=0.7}[][]{\footnotesize$\:\:\:\:\:\:\rho=0.7$}
\psfrag{rho=1}[][]{\footnotesize$\:\:\:\:\:\:\rho=1$}
\psfrag{unquantized system, rho=0.1}[][]{\footnotesize unquantized, $\rho\!=\!0.1$}
\psfrag{l1=l2=4}[][]{\footnotesize${\ell}_u={\ell}_d=4$}
\psfrag{deltan}[][]{\footnotesize$\Delta n$}
\psfrag{Required SNR @BER=10}[][]{\footnotesize Required SNR @BER$=10^{-3}$ in dB}
\epsfig{file= 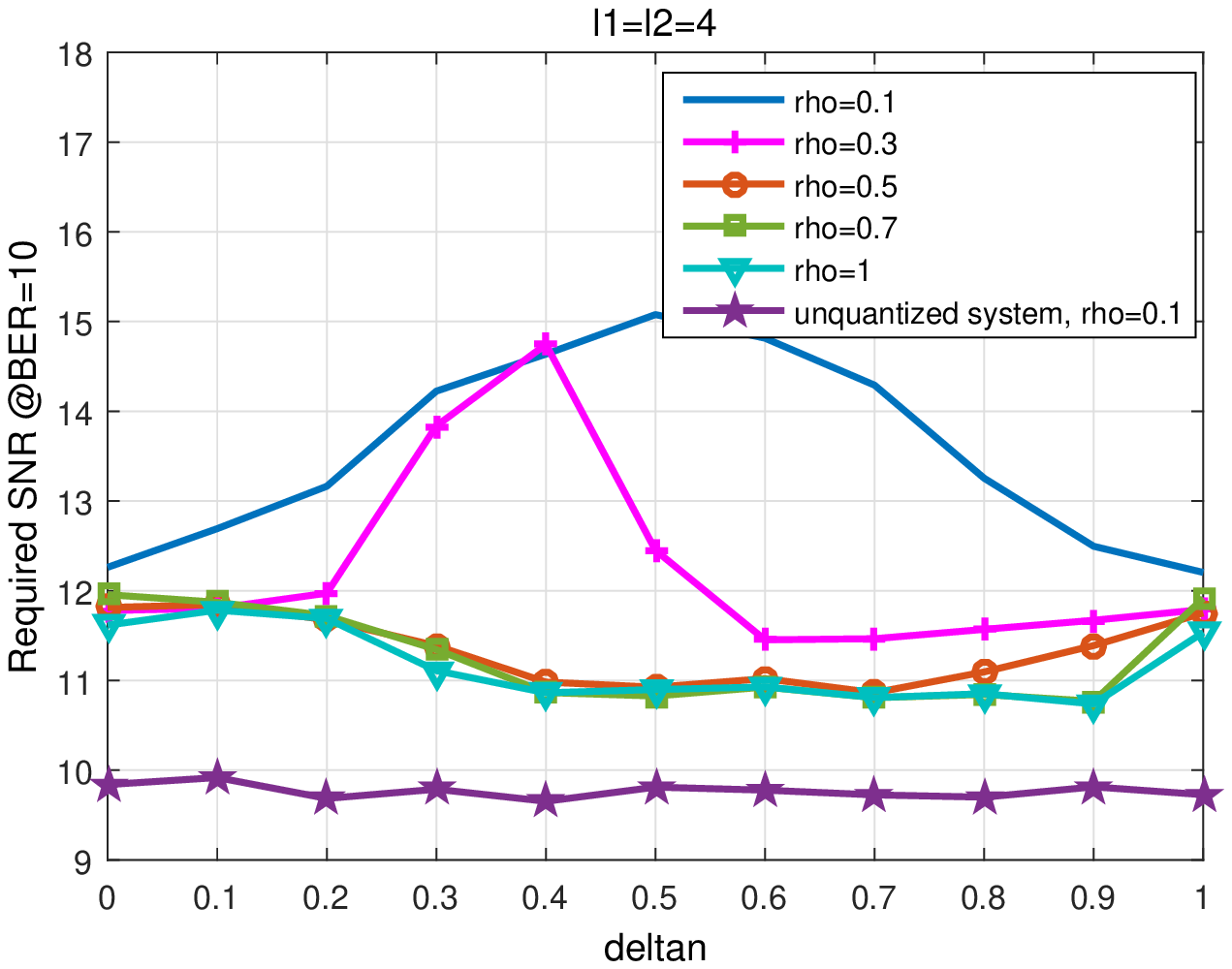, width = \columnwidth}
\end{minipage}
\begin{minipage}{\columnwidth}
\psfrag{rho=0.1}[][]{\footnotesize $\:\:\:\:\:\:\rho=0.1$}
\psfrag{rho=0.3}[][]{\footnotesize$\:\:\:\:\:\:\rho=0.3$}
\psfrag{rho=0.5}[][]{\footnotesize$\:\:\:\:\:\:\rho=0.5$}
\psfrag{rho=0.7}[][]{\footnotesize$\:\:\:\:\:\:\rho=0.7$}
\psfrag{rho=1}[][]{\footnotesize$\:\:\:\:\:\:\rho=1$}
\psfrag{unquantized system, rho=0.1}[][]{\footnotesize  unquantized,\! $\rho\!=\!0.1$}
\psfrag{l1=l2=4}[][]{\footnotesize${\ell}_u=4$}
\psfrag{deltan}[][]{\footnotesize$\Delta n$}
\psfrag{B}[][]{\footnotesize $B_{0.9375}$ in $1/T_s$}
\epsfig{file= 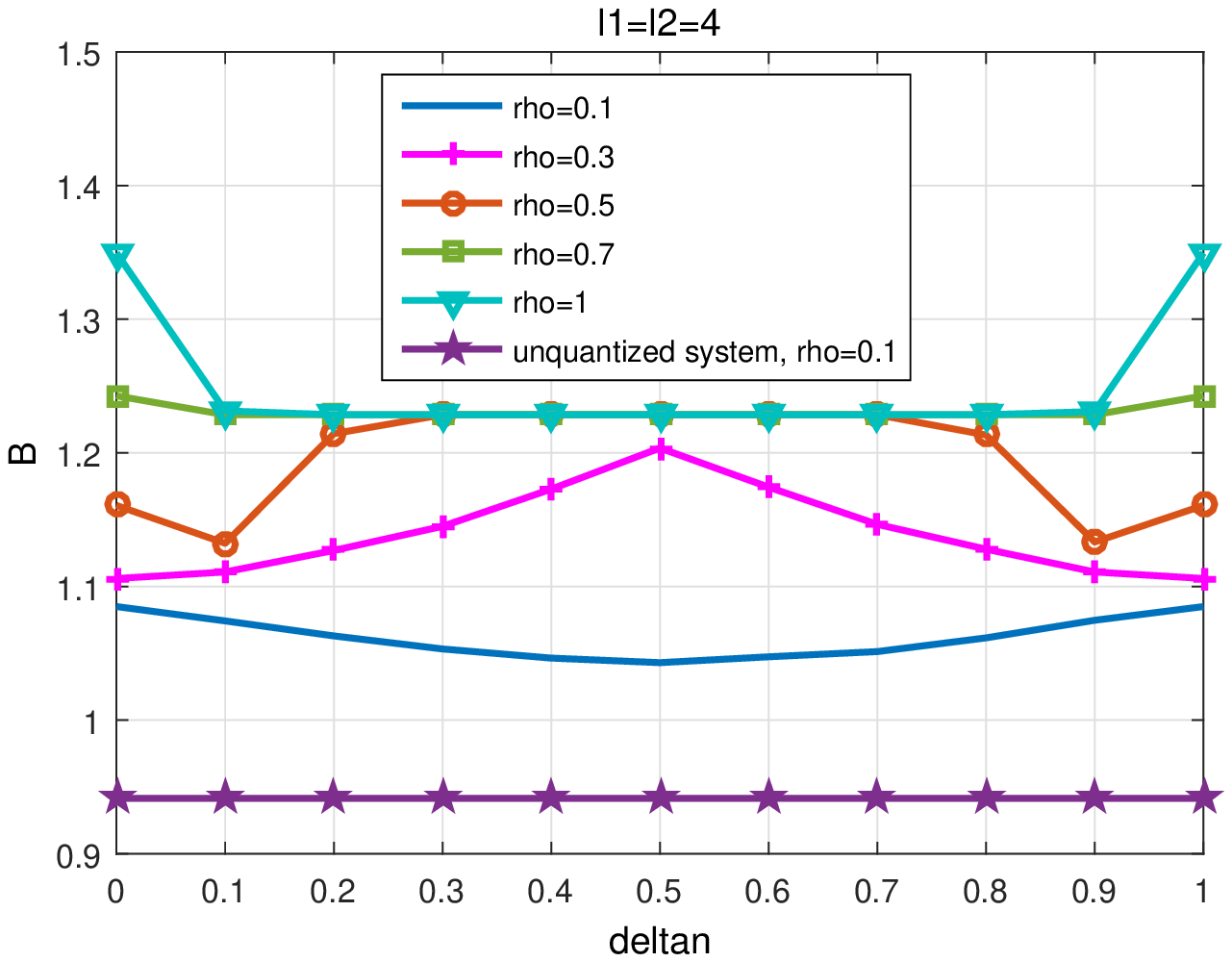, width = \columnwidth}
\end{minipage}
\caption{Required SNR @BER of $10^{-3}$ and $93,75\%$ bandwidth as function of the fractional delay $\Delta n$ for different roll-off factors $\rho$ and for $l_u\!=\!l_d\!=\!4$.}
\label{fig:results_vs_dn_l_4}
\end{figure}

\begin{figure}[h]
\centering  
\begin{minipage}{\columnwidth}
\psfrag{rho = 0.1}[][]{\footnotesize \:\:\:\:\:\: $\rho=0.1$}
\psfrag{rho = 0.3}[][]{\footnotesize$\:\:\:\:\:\:\rho=0.3$}
\psfrag{rho = 0.5}[][]{\footnotesize$\:\:\:\:\:\:\rho=0.5$}
\psfrag{rho = 0.7}[][]{\footnotesize$\:\:\:\:\:\:\rho=0.7$}
\psfrag{rho = 1}[][]{\footnotesize$\:\:\:\:\:\:\rho=1$}
\psfrag{unquantized system, rho=0.1}[][]{\footnotesize \: unquantized,\! $\rho\!=\!0.1$}
\psfrag{l1=l2=8}[][]{\footnotesize${\ell}_u={\ell}_d=8$}
\psfrag{deltan}[][]{\footnotesize$\Delta n$}
\psfrag{Required SNR @BER=10}[][]{\footnotesize Required SNR @BER$=10^{-3}$ in dB}
\epsfig{file= 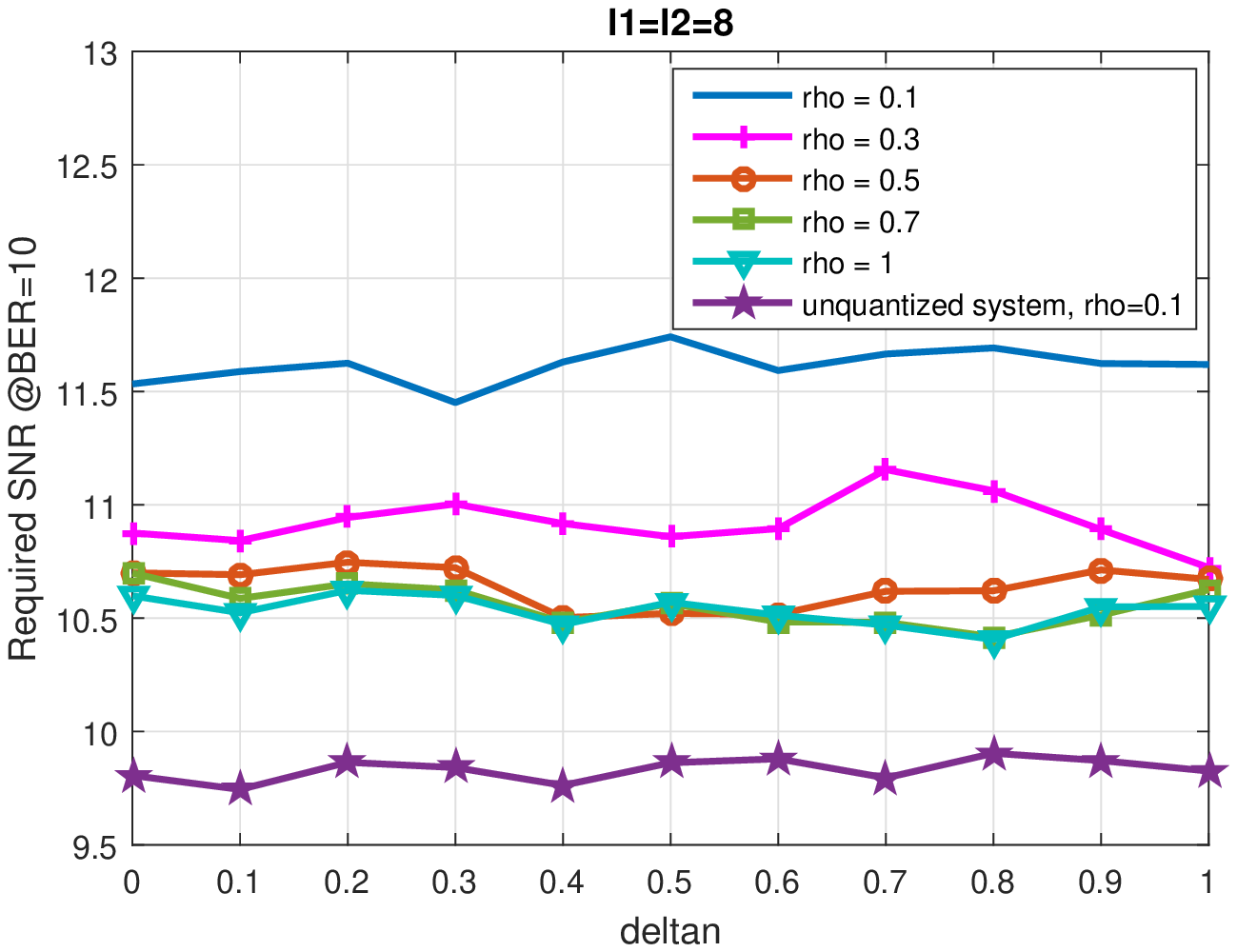, width = \columnwidth}
\end{minipage}
\begin{minipage}{\columnwidth}
\psfrag{rho = 0.1}[][]{\footnotesize $\:\:\:\:\:\:\rho=0.1$}
\psfrag{rho = 0.3}[][]{\footnotesize$\:\:\:\:\:\:\rho=0.3$}
\psfrag{rho = 0.5}[][]{\footnotesize$\:\:\:\:\:\:\rho=0.5$}
\psfrag{rho = 0.7}[][]{\footnotesize$\:\:\:\:\:\:\rho=0.7$}
\psfrag{rho = 1}[][]{\footnotesize$\:\:\:\:\:\:\rho=1$}
\psfrag{unquantized system, rho=0.1}[][]{\footnotesize \: unquantized,\! $\rho\!=\!0.1$}
\psfrag{l1=8}[][]{\footnotesize${\ell}_u=8$}
\psfrag{deltan}[][]{\footnotesize$\Delta n$}
\psfrag{B}[][]{\footnotesize $B_{0.9375}$ in $1/T_s$}
\epsfig{file= 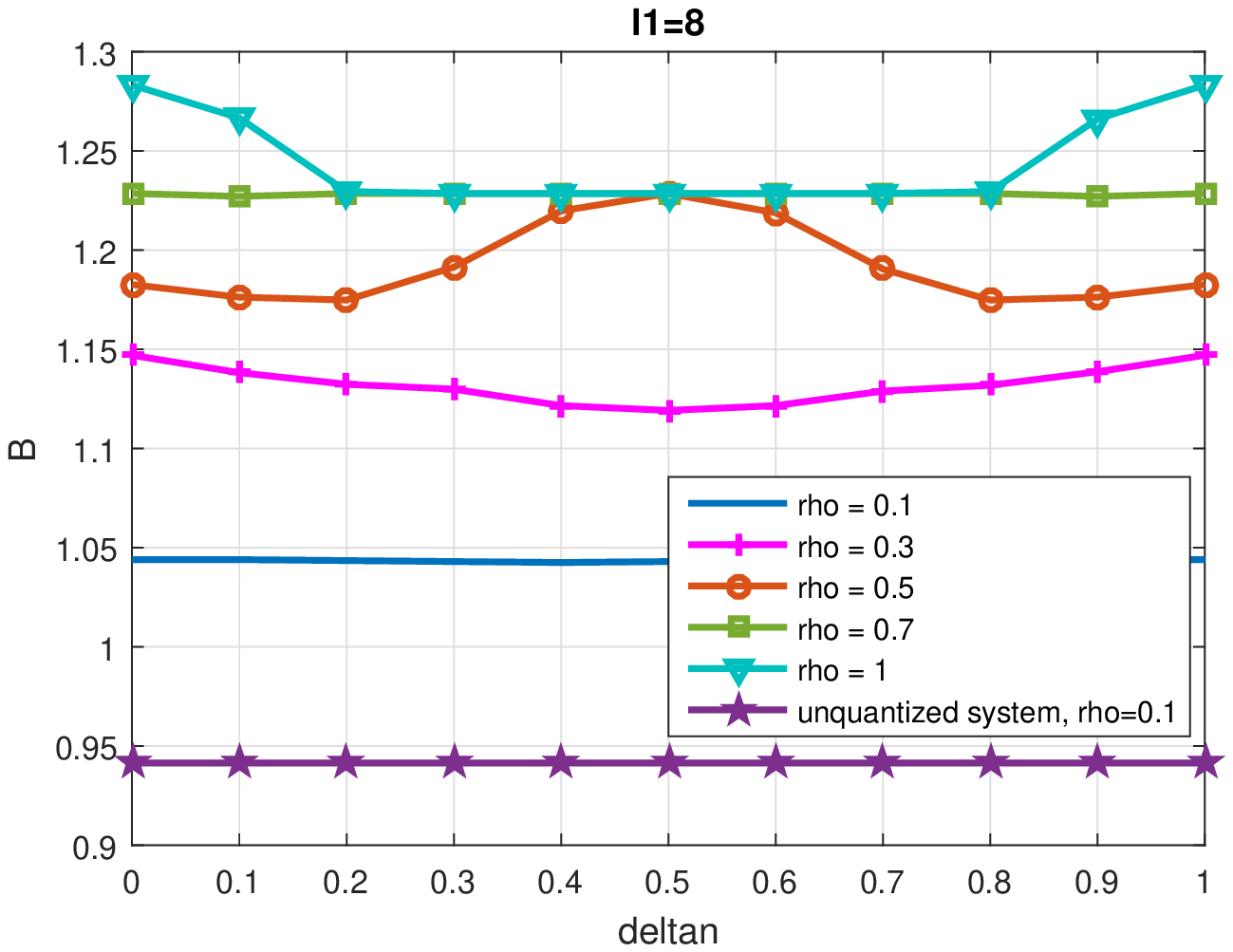, width = \columnwidth}
\end{minipage}
\caption{Required SNR @BER of $10^{-3}$ and $93,75\%$ bandwidth as function of the fractional delay $\Delta n$ for different roll-off factors $\rho$ and for $l_u\!=\!l_d\!=\!2$.}
\label{fig:results_vs_dn_l_8}
\end{figure}

\begin{figure}[h]
\centering  
\psfrag{l2 = 2}[][]{\:\:\:\:\:\:\:\:\:\:\:\:\:\:\:\: \footnotesize${\ell}_u=2,$ ${\ell}_d=2$}
\psfrag{l2 = 4}[][]{\:\:\:\:\:\:\:\:\:\:\:\:\:\:\:\:\:\footnotesize${\ell}_u=2,$ ${\ell}_d=4$}
\psfrag{l2 = 8}[][]{\:\:\:\:\:\:\:\:\:\:\:\:\:\:\:\: \footnotesize${\ell}_u=2,$ ${\ell}_d=8$}
\psfrag{l1 = l2 = 4 to compare}[][]{\!\!\!\!\!  \footnotesize${\ell}_u={\ell}_d=4$}
\psfrag{l1 = l2 = 8}[][]{ \:\:\:\:\:\:\: \footnotesize${\ell}_u={\ell}_d=8$}
\psfrag{deltan}[][]{\footnotesize$\Delta n$}
\psfrag{Required SNR @BER=10}[][]{\footnotesize Required SNR @BER$=10^{-3}$ in dB}
\psfrag{rho=0.1, l1=2}[][]{\footnotesize$\rho=0.1$}
\epsfig{file= 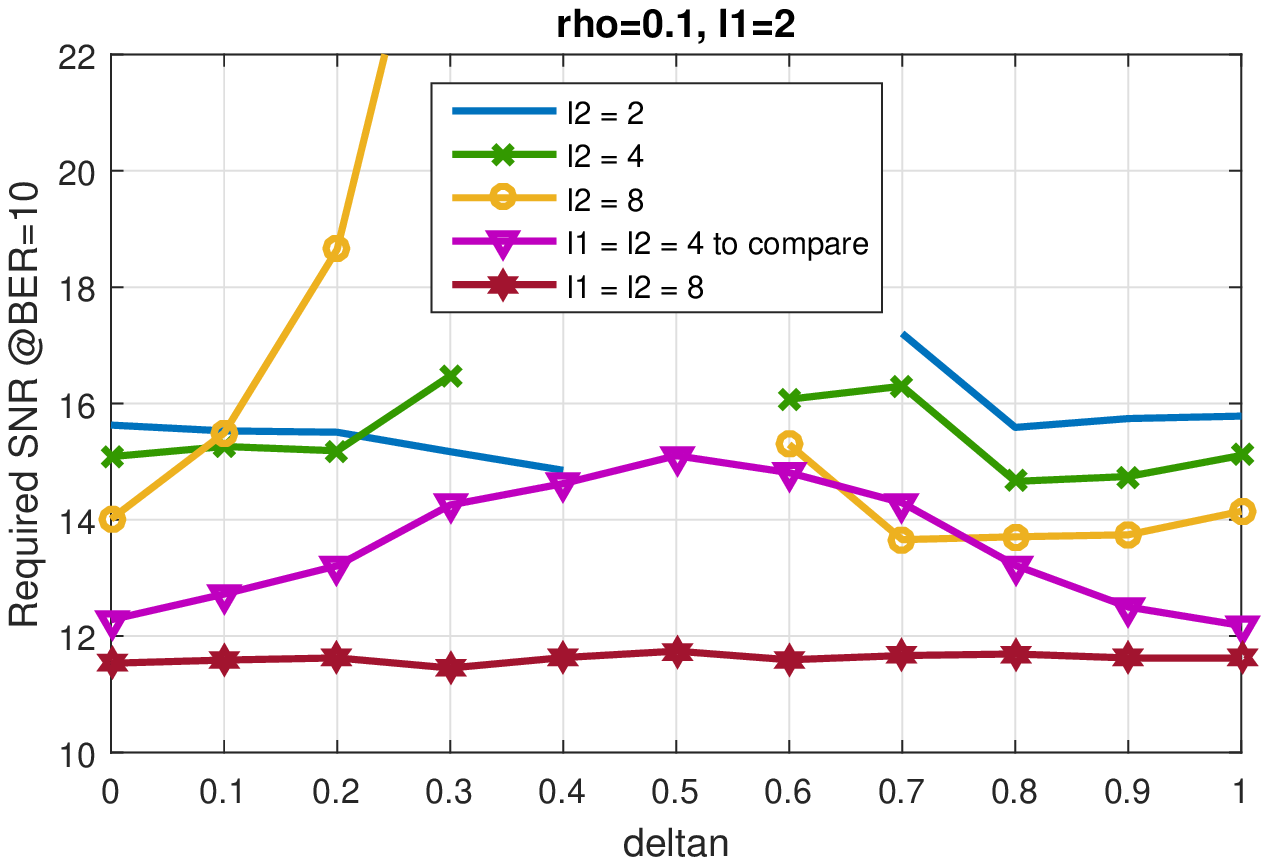, width = \columnwidth}
\caption{Required SNR @BER of $10^{-3}$ for $\rho=0.1$ and different oversampling factors at the transmitter and the receiver.}
\label{fig:results_vs_l2}
\end{figure}

In Fig.~\ref{fig:results_vs_dn_l_4}, we increase the oversampling factors to $l_u=l_d=4$. In this case, the performance becomes less sensitive to the choice of $\Delta n$. The improvements in terms of bandwidth as well as required SNR are more obvious for small roll-off values. Similar conclusions can be drawn in Fig.~\ref{fig:results_vs_dn_l_8} when increasing the oversampling factors to $l_u=l_d=8$, where the performance gets very close to the ideal case.  For better understanding of the influence of higher oversampling, we consider in Fig.~\ref{fig:results_vs_l2} the case of $\rho=0.1$ while increasing $l_d \in \{4,8\}$ at receiver and keeping the transmit oversampling factor $l_u=2$ on the one hand, and on the other hand we increase both oversampling factors simultaneously $l_u=l_d \in \{ 2,4,8 \}$ at both ends of the communication link. It can be seen that the best SNR performance can be achieved with simultaneous oversampling $l_u=l_d=8$ regardless of the fractional delay $\Delta_n$. This result suggests that oversampling at both sides of the communication link is  beneficial in terms of SNR performance for the 1-bit system, which can be explained by the fact that the quantization error is spread over higher bandwidth reducing the total noise power in the desired band.

\subsection{Spectral shape}
\begin{figure}[h]
\centering  
\psfrag{rho=0.1, l1======2}[][]{\footnotesize  $\rho=0.1$, ${\ell}_u\!\!=\!\!2$}
\psfrag{rho=0.1, l1======4}[][]{\footnotesize  $\rho=0.1$, ${\ell}_u\!\!=\!\!8$}
\psfrag{rho=1, l1======2}[][]{\footnotesize $\rho=1$, ${\ell}_u\!\!=\!\!2$}
\psfrag{rho=1, l1======4}[][]{\footnotesize $\rho=1$, ${\ell}_u\!\!=\!\!8$}
\psfrag{PSD in dB}[][]{\footnotesize PSD in dB}
\psfrag{Frequency}[][]{\footnotesize Frequency in $1/T_s$}
\epsfig{file= 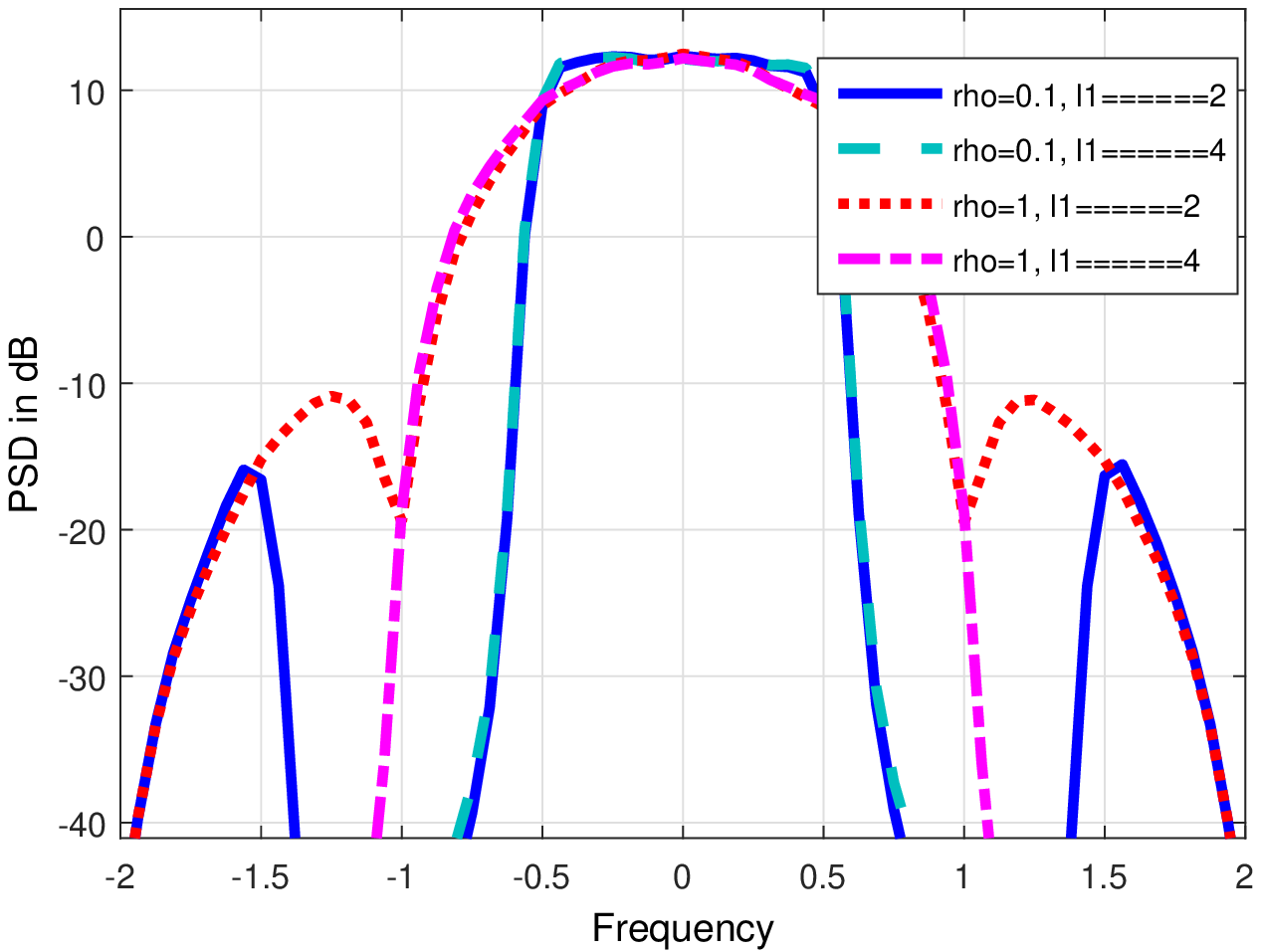, width = \columnwidth}
\caption{PSD of the unquantized system}
\label{fig:psd}
\end{figure}

\begin{figure}[h]
\centering  
\psfrag{rho=0.1, l1=l2=2, dn==========0.4}[][]{\footnotesize  $\rho=0.1$, ${\ell}_u\!\!=\!\!2$, $\Delta n\!\! =\!\!0.4$}
\psfrag{rho=0.1, l1=l2=8, dn=0}[][]{\:\:\:\:\:\:\:\:\:\:\:\:\:\footnotesize $\rho=0.1$, ${\ell}_u\!\!=\!\!8$, $\Delta n\!\! =\!\!0$}
\psfrag{rho=1, l1=l2=2, dn=0.5}[][]{\:\:\:\:\:\:\:\:\:\:\:\:\:\footnotesize $\rho=1$, ${\ell}_u\!\!=\!\!2$, $\Delta n \!\!=\!\!0.5$}
\psfrag{rho=1, l1=l2=8, dn=0.5}[][]{\:\:\:\:\:\:\:\:\:\:\:\:\footnotesize $\rho=1$, ${\ell}_u\!\!=\!\!8$, $\Delta n\!\! =\!\!0.5$}
\psfrag{PSD in dB}[][]{\footnotesize PSD in dB}
\psfrag{Frequency}[][]{\footnotesize Frequency in $1/T_s$}
\epsfig{file= 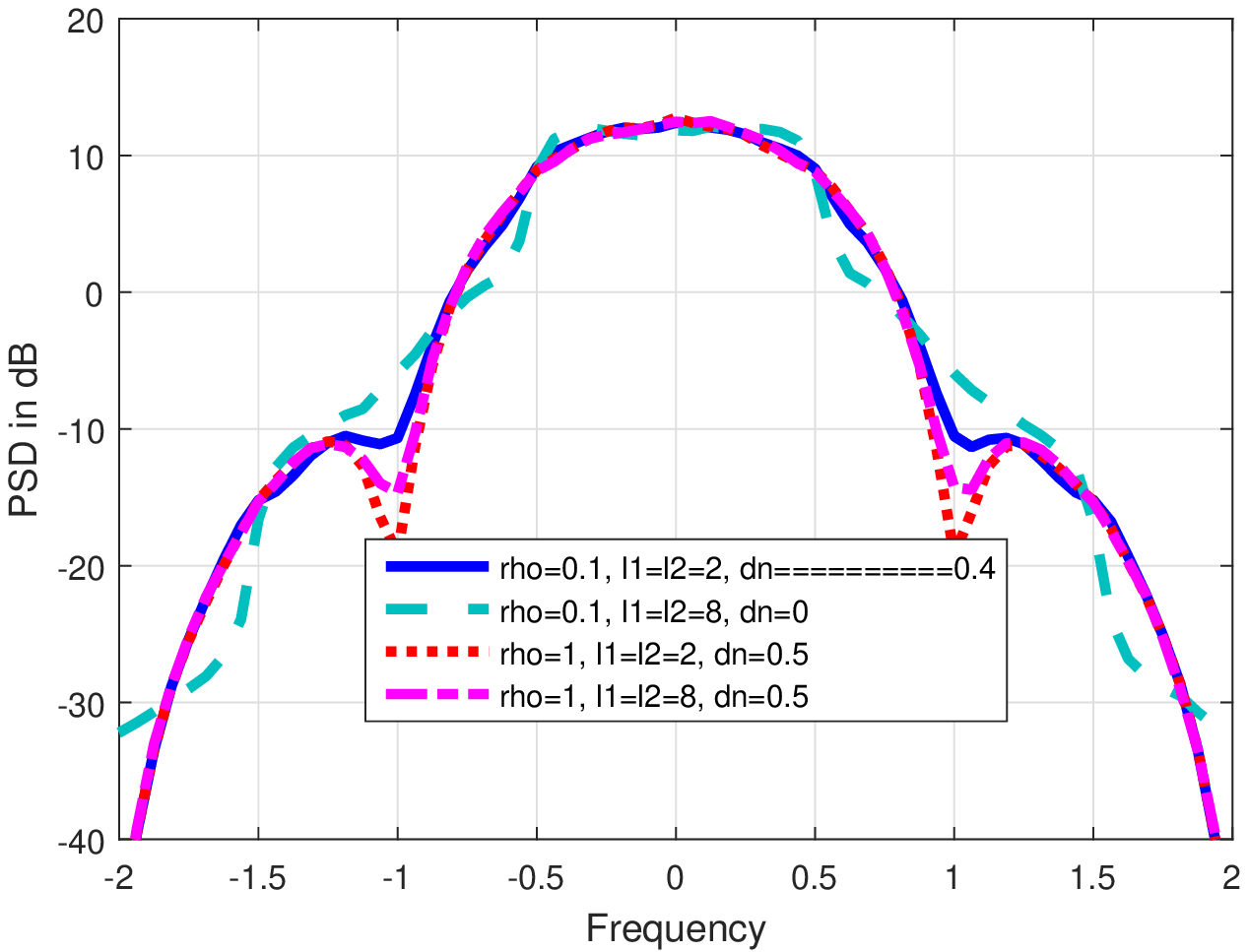, width = \columnwidth}
\caption{PSD of the quantized system}
\label{fig:psd_quantized}
\end{figure}

For illustration, Fig.~\ref{fig:psd_quantized} shows the power spectral density  (PSD) of the 1-bit quantized signal after the LPF for different roll-off factors, while Fig.~\ref{fig:psd}  depicts the corresponding PSD without quantization. Contrarily to the unquantized case, where oversampling can help relaxing the demands on the LPF, the spectral shape of the 1-bit system does not benefit from the oversampling. In fact, the spectral shape at the pass band region is mainly formed by the digital RRC filter, while the stop band is mainly influenced by the LPF. Thus it does not improve with higher oversampling factors.

\section{Conclusion}

Low resolution ADCs and DACs  are very advantageous in terms of  the system complexity especially in the context of massive MIMO. Here, we considered a low cost single carrier communication system, where 1-bit DAC is applied to RRC spectrally shaped signals and 1-bit ADC is used at the receiver side.  By tuning the fractional delay of the RRC impulse response and modifying the linear receiver to take into account the effects of coarse quantization,  i.e. without any extra complexity, we have shown that the performance in terms of spectral efficiency and radiated power efficiency can be made quite close to the ideal system especially for moderate roll-off values. In addition, we have shown that oversampling beyond factor two is still beneficial  especially for small roll-off values.

\bibliographystyle{IEEEbib}
\bibliography{IEEEabrv,references}
\end{document}